\documentclass{elsarticle}
\usepackage[utf8]{inputenc}
\usepackage{microtype}
\usepackage{lmodern}
\usepackage{amsmath, amsfonts,amsthm,amssymb,color}
\usepackage{titlesec}
\usepackage{graphicx}

\theoremstyle{plain}

\theoremstyle{definition}
\newtheorem{definition}{Definition}

\theoremstyle{remark}

\begin{document}

\begin{frontmatter}
	\title{Best Portfolio Management Strategies For Synthetic and Real Assets}
	
	\author[1]{Jarosław Gruszka\corref{cor1}}
	\ead{jaroslaw.gruszka@pwr.edu.pl}
	
	\author[1]{Janusz Szwabi\'{n}ski}
	\ead{janusz.szwabinski@pwr.edu.pl}
	
	\cortext[cor1]{Corresponding author}
	\address[1]{Hugo Steinhaus Center, Faculty of Pure and Applied Mathematics, Wrocław University of Science and Technology}
	
	\begin{abstract}
		Managing investment portfolios is an old and well know problem in multiple fields including financial mathematics and financial engineering as well as econometrics and econophysics. Multiple different concepts and theories were used so far to describe methods of handling with financial assets, including differential equations, stochastic calculus and advanced statistics. In this paper, using a set of tools from the probability theory, various strategies of building financial portfolios are analysed in different market conditions. A special attention is given to several realisations of a so called balanced portfolio, which is rooted in the natural ``buy-low-sell-high'' principle. Results show that there is no universal strategy, because they perform differently in different circumstances (e.g. for varying transaction costs). Moreover, the planned time of investment may also have a significant impact on the profitability of certain strategies. All methods have been tested with both simulated trajectories and real data from the Polish stock market.
	\end{abstract}



\end{frontmatter}

\section{Introduction}

The interest in selecting and holding investment assets, which we nowadays refer to as building and managing investment portfolio is an old and well know problem in the field of many disciplines including financial mathematics, financial engineering and econometric. The foundation of the theory of selecting assets for an investment portfolio was laid by an American economist and Noble Prize winner -- Harry Markowitz. In his article from 1952 he postulated that an investor acting rationally should always try to maximise their expected returns and/or minimise the risk of the entire portfolio, taking into account possible correlations between the assets~\cite{markowitz_1952}. The approach presented in the article turned out to be extremely well-received and the strategies described there have been collectively named a Modern Portfolio Theory or MTP in short. The concept introduced by Markowitz was then widely developed and built up. In 1963 William Sharpe devised a very computationally efficient method of analysing available assets and creating a portfolio having desired properties \cite{sharpe_1963} and in 1972 Robert C. Merton published a paper in which he shows how to analytically obtain the "efficient frontier" --  a  graphical illustration of a set of all portfolios optimal in Markowitz sense \cite{merton_1972}. Notable extensions of Markowitz's model were made by G.A. Pogue who enhanced the model by taking into consideration transactions' costs and short selling \cite{pogue_1970}. In 1992 Fisher Black and Robert Litterman built their own model based on the one of Markowitz, which did not require to associate assets with specific expected returns (which were difficult to estimate) \cite{black_1992} and in 1991 B.M. Rom and K. Ferguson extended that model even further by addressing the limitations of the original model regarding the distribution of returns and the usage of variance of returns as a measure of investment's risk \cite{rom_1994}. What they accomplished is often called a Post Modern Portfolio Theory. More recent research is also still being conducted within the Markowitz framework and newest modelling and problem-solving approaches are applied. In 2013 M. Woodside-Oriakhi et.al. presented an extensive model of portfolio optimization and solved the stated problem by mixed-integer quadratic programming \cite{woodside-oriakhi_2013}. In 2014 Mittal and Mehlawat proposed a model including transaction costs and also solved an associated optimization problem but by means of real-coded genetic algorithms \cite{mittal_2014}.

There were also numerous publications on the topic of portfolio management where the fast developing theory of fuzzy sets was applied. One of the pioneering works in that field was the one of Hideo Tanakaa et.al. from 1998 in which authors presented the first extension of Markovitz's idea to use the fuzzy set theory and exchanging the notion of probability for, what is called, fuzzy probability \cite{tanaka_2000}. Research was later conducted by i.a. Christer Carlsson et.al. who were successful in finding an optimal portfolio in Markowitz sense assuming that returns are fuzzy variables (instead of classical random variables) \cite{carlsson_2002}. Also, Yong Fang et.al. introduced the idea of portfolio rebalancing within the fuzzy set theory framework \cite{fang_2006}. By rebalancing we understand changing portfolio composition after some time from the beginning of an investment, reason being usually investor's willingness to adjust to the market movements which happened in the meantime. The concept of rebalancing will be further discussed and frequently used in the latter part of the paper.

The idea of changing the investment portfolio in time was actually devised independently of the Markowitz framework and Paul A. Samuelson was one of the firsts to study it. In his article from 1969 he was considering the problem how to rebalance a life-time portfolio at the number of discrete
moments of time \cite{samuelson_1969}. Similar problems were then studied by Robert C. Merton in his two famous articles from 1969 and 1971 in which he discusses the topic of building a life-time portfolio of stocks for the sake of calculating optimal consumption (amount of money that can be withdrawn from portfolio's value with no major detriment for this portfolio's performance) \cite{merton_1969, merton_1971}. The other worth-noticing pieces of work in this stream were i.a. the paper of Morton and Pliska in which they introduced transaction costs to the Merton portfolio problem and eliminated the need of making transactions in continuous time, giving instead a sequence of discrete moments of time at which transactions are allowed to be made by an investor \cite{morton_1995}. There also was a paper by Colin Atkinson et.al. in which the portfolio problem in question was translated into the language of differential equations and a numerical procedure to solve them was proposed \cite{atkinson_1997}. On the other hand, Jak\v{s}a Cvitani\'c  and Ioannis Karatzas were studying a slightly different portfolio optimization problem using martingale theory and advanced tools of stochastic calculus \cite{cvitanic_1996}.

An entirely different approach to the topic of portfolio optimisation has been also proposed in regard to a famous paper of J.L. Kelly Jr \cite{kelly_1956}. The paper itself is not however related to portfolio management directly -- actually all statements are formulated in the language of information theory and examples given are referencing gambling. The author argues that the best strategy for a gambler is to maximise the logarithm of wealth (which in this context simply meant the logarithm of the amount of money possessed by the gambler). Kelly proved that a strategy devised that way will outperform any other strategy in a long enough amount of time almost surely. This idea was called a 'Kelly criterion' and it was quickly implemented in the field of portfolio management and found a lot of attention but also a lot of controversy there. One of the biggest criticisers of this approach were Paul Samuelson and Robert C. Merton, who were of the opinion that the research related to portfolio management should continue to be conducted in the language of utility function (developed by them) calling the application of 'Kelly criterion' a 'fallacy' \cite{samuelson_1971}. Despite this criticism, the idea of building and maintaining a Growth Optimal Portfolio (which evolved from the 'Kelly criterion') is still being extensively studied. The first chapter of a book by Morten Mosegaard Christensen from 2012 constitutes a very broad overview of the history, mathematical and practical aspects and various discussions on the topic of the Growth Optimal Portfolio \cite{christensen_2012}. A study of a portfolio using Kelly strategy has also been presented in an article of Paolo Laureti et.al. \cite{laureti_2010}.

As presented above, various methodologies, concepts and ideas have been utilised to study portfolio management. One of the techniques which we have not mentioned earlier while it is getting increasingly popular recent times are Monte Carlo simulations. They were used by Riccardo Cesari and David Cremonini who provided a wide comparison between various portfolio management techniques in their article from 2003 \cite{cesari_2003}. Their main assumption was the normality and independence of returns which they were first testing in the data and later simulating. A slightly more theoretically-oriented article written by Ofer Alper et.al. appeared in 2017 \cite{alper_2017} aiming to compare selected portfolio management strategies in the presence of correlation between the assets and transaction fees. 

This work can be seen as a continuation of the research presented in Alper et.al \cite{alper_2017}. We disambiguate notions and ideas presented there. Moreover, we present the new outcomes resulting from our own methodology of testing various portfolio management strategies. Instead of selecting few instruments and demonstrating the effects of some management strategies on a particular portfolio created this way, we devise a method of randomly choosing multiple different portfolios and averaging results for all of them. Such approach brings more objectivity to the comparison between various strategies.

\section{Model description}

\subsection{Principles of portfolio dynamics}
As a first step, we are going to formally define the notions that we will be using throughout the article. We begin with \emph{a market}. 

\begin{definition}
	\label{definition_market}
	A market $\mathcal{M}$ is a set of $N$ stochastic processes, $N \in \mathbb{N}\setminus \{0\}$, defined on the same time domain $[0, T]$.
\end{definition}

Each of the stochastic processes in $\mathcal{M}$ is going to represent the price of a single asset traded on some stock exchange market over the time from $0$ to $T$, measured in years.

We can proceed with the definition of a \emph{portfolio} on a market. 

\begin{definition}
	\label{definition_portfolio}
	A portfolio $\mathcal{P}$ built on a market $\mathcal{M}$ is an ordered pair $(S(t), Q(t))$. $S$ is called an asset component and it is a stochastic, $n$-dimensional vector-valued function, $n \in \mathbb{N}\setminus \{0\}, n \leq N$ with coordinate functions being elements of $\mathcal{M}$. In other words, $S$ is of the form $S(t) = (S_1(t), S_2(t), \ldots, S_n(t))$ and for all $i \in \{1,2, \ldots , n\}, S_i \in \mathcal{M}$. $Q$ is called a quantity component and it is a deterministic $n$-dimensional vector-valued function with coordinate functions $q_i(t): [0, T] \rightarrow \mathbb{R}$ for $i \in \{1, 2, \ldots , n\}$. 
\end{definition}

The elements of the quantity component at any given moment of time represent the amount of a given asset in a portfolio, e.g. number of shares if the assets we are considering is common stock.

To evaluate portfolio performance, number of measures can be introduced. The most basic one is its value changing over time. We call it portfolio \emph{wealth}.

\begin{definition}
	\label{definition_wealth}
	Wealth of portfolio $\mathcal{P}$ at any moment of time $t\in [0,T]$ is a scalar product of the vector of values of assets $S$ and the vector of quantities of these assets~$Q$.
	\begin{equation}
	\label{formula_wealth}
	W(t) = \langle S(t), Q(t) \rangle = \sum_{i=1}^{n}{S_i(t)q_i(t)},
	\end{equation}
\end{definition}

Although better measures will be introduced at a later stage, we will use this one to describe portfolio management strategies that we are going to study. 

\subsection{Portfolio management strategies}

The first and the simplest strategy is associated with building what is called \emph{a passive portfolio}. It is a one in which the quantities of assets are fixed at the initial moment of time $t = 0$ and are not being altered during the entire time of portfolio lifetime.

\begin{definition}
	\label{definition_passive_portfolio}
	A passive portfolio $\mathcal{P}_{pass}$ is a portfolio for which values of the functions of the quantity component are constant in time, i.e. for each $q_i(t)$ we have $q_i(t) = q_i(0) = const$ for any $t$.
\end{definition}

This method of conducting an investment portfolio is also sometimes referred to as \emph{a buy-and-forget portfolio}. It is admittedly very straightforward but it can be treated as kind of a benchmark -- to check if reacting for the market movements in some specific way can lead to a significant improvement of portfolio performance. 

A slightly more complex method of portfolio management, which is based on undertaking specific actions when market prices change, is called \emph{a balanced portfolio}. To define this method, we first need to define the concept of \emph{fractions of wealth}.

\begin{definition}
	\label{definition_wealth_fraction}
	An $i$-th fraction of wealth $f_i$ (also called a wealth fraction or simply -- a fraction), at any moment of time $t$, is a component of the portfolio wealth $W(t)$ associated with an $i$-th portfolio asset, i.e.
	\begin{equation}
	\label{formula_fractions}
	f_i(t) = \frac{S_i(t)\cdot q_i(t)}{W(t)} \quad \text{for all } i \in \{1,2 \ldots n\}.
	\end{equation}
\end{definition}

Knowing the meaning of wealth fractions, we can define a balanced portfolio.

\begin{definition}
	\label{definition_balanced_portfolio}
	A balanced portfolio $\mathcal{P}_{bal}$ is a portfolio for which all fractions of wealth are constant in time, i.e. for each for all  $i \in \{1,2 \ldots n\}$ we have $f_i(t) = f_i(0) = const$ for any $t$.
\end{definition}

Such a portfolio construct essentially guarantees that if price of a particular asset increases, its amount in the portfolio is made smaller (hence -- assets are sold when their prices are getting higher) and if the price drops -- its amount is increased (thus -- we buy new assets when their prices are diminishing). However, the above definition does not provide any guidance on how to manipulate the quantities of the assets to keep portfolio in a balanced state. In fact, it is not easy to show such procedure in a general case, i.e. for any moment of time $t$, especially if $t$ changes continuously. However, this is not actually necessary as in practice we hardly ever deal with continuous pricing of assets and even if we were -- we would be unable to perform market transactions in a continuous manner. Therefore, for the sake of computer simulations as well as for the practical application --  it is sufficient to create an iterative procedure which captures the idea of building a balanced portfolio. To this end we introduce the sequence of discrete moments of time, equally distributed every $\Delta t$, i.e. $t_0 = 0, t_1 = \Delta t, t_2 = 2 \Delta t, t_3 = 3 \Delta t \ldots$ where $0<\Delta t \ll T$. We then fix the initial number of each kind of assets $(q_1(0), q_2(0), \ldots, q_n(0))$, and based on that we compute the initial wealth of portfolio according to the formula \eqref{formula_wealth} and the wealth fractions as per the formula \eqref{formula_fractions} and we compute the quantities of assets in the following way 

\begin{equation}
\label{formula_updating_quantities}
q_i(k\Delta t) = f_i \frac{W^{temp}(k\Delta t)}{S_i(k\Delta t)} 
\end{equation} 
for consecutive $k, k \in \mathbb{N}, 1\leqslant k \leqslant \frac{T}{\Delta t}$. The factor $W^{temp}(t)$ can be called the temporary wealth of portfolio to distinguish it to the ordinary (ultimate) portfolio wealth $W(t)$ given by formula \eqref{formula_wealth}. Temporary portfolio wealth is calculated based on the fact that before the rebalance is done, for each asset its quantity is equal to the actual quantity of this asset from previous step. Therefore $W^{temp}(k\Delta t) = \sum_{i=1}^{n}{S_i(t)q^{temp}_i(t)}$ where $q^{temp}_i(k\Delta t) = q_i((k-1)\Delta t)$. Only once $q_i(t)$ is computed (which we can think of as making the actual rebalancing transactions) one can calculate the true (ultimate) wealth of portfolio $W(t)$, using the formula \eqref{formula_wealth}.

Since in this paper we will only be working with the discrete time assumption, from that moment onward, whenever we will be speaking about a 'balanced portfolio' we will be referring to a construct described by the quantity update formula \eqref{formula_updating_quantities}, not the general concept presented in Definition \ref{definition_balanced_portfolio}. 

It is also worth to notice that a good portfolio management strategy should be self-financing i.e. apart from the initial amount of cash needed to purchase the assets at $t=0$, no additional cash inflow should be needed to keep portfolio in a balanced state as well as all portfolio profits should immediately get reinvested. The updating procedure described by equation \eqref{formula_updating_quantities} which in some sense defines the entire strategy, also makes it meet the expectation of being self-financing. To prove it, let us get a closer look at the wealth of portfolio at an arbitrary point of time $k\Delta t$. Its value can be obtained by using formulas \eqref{formula_wealth} and \eqref{formula_updating_quantities}. We have

\begin{equation*}
\begin{split}
W(k\Delta t) =  
\sum_{i=1}^{n}{S_i(k\Delta t)q_i(k\Delta t)} &=
\sum_{i=1}^{n}{S_i(k\Delta t)f_i \frac{W^{temp}(k\Delta t)}{S_i(k\Delta t)}} \\
&= \sum_{i=1}^{n}{f_i W^{temp}(k\Delta t)} = 
W^{temp}(k\Delta t)\sum_{i=1}^{n}{f_i}.
\end{split}
\end{equation*}

From the Definition \ref{definition_wealth_fraction} of wealth fractions we know that $\sum_{i=1}^{n}{f_i} = 1$. Henceforth, we have 

\begin{equation*}
W(k\Delta t) = W^{temp}(k\Delta t)
\end{equation*}
which indicates that performing a rebalance operation does not require any additional money and all changes in portfolio wealth are results of fluctuating assets prices -- there is neither any kind of withdrawal nor insertion of money from or into the portfolio.  

\subsection{Transactions costs}

Most markets charge investors for making transactions, which is why keeping portfolio balanced becomes expensive if we take transaction fees into consideration. In most cases, the value of the fees is determined as a percentage of the value of assets being exchanged, which effectively lowers the amount of money one can use to reconstruct the portfolio and make it balanced again. If we denote this percentage by $\alpha$, the value of assets of $i$-th type which is exchanged by $V_i(t)$ and the cumulative value of fees by $F(t)$ then the dependencies between these quantities are given by the following formulas:

\begin{align}
V_i(t) &= S_i(t) \cdot (q^{temp}_i(t) - q_i(t)) \label{formula_vale_of_exchanged_assets} \\
F(t) &= \sum_{i=1}^{n}{\alpha|V_i(t)|} \label{formula_fees}.
\end{align}

Following the notation presented above, we can derive the formula for updating quantities of assets in case of a balanced portfolio

\begin{align}
q_i(k\Delta t) &= f_i \frac{W^{temp}(k\Delta t) - F(k\Delta t)}{S_i(k \Delta t)}. \label{formula_updating_quantities_fees}
\end{align} 

Note that if we take transaction costs into consideration, as described by formula \eqref{formula_updating_quantities_fees}, keeping portfolio in a balanced state becomes costly so that profits resulting from using this strategy might be overtaken by the expenses related to the fees. This is why two more frugal strategies have been devised. First one is based on the idea of rarefying the moments of rebalancing the portfolio, by some constant $m, m~\in~\mathbb{N}~\setminus~\{0, 1\}$, i.e. performing the rebalance $m$ times less often. This strategy is called a \emph{periodically balanced portfolio}. In such case the procedure of updating the quantities of assets takes place once every $m$ time intervals of length $\Delta t$ and between these moments portfolio acts like a passive one. This procedure can be summarised by the following formula:

\begin{align}
\label{formula_updating_quantities_periodic_rebalance_fees}
q_i(k\Delta t) &= 
\left\{
\begin{array}{ll}
f_i \frac{W^{temp}(k\Delta t) - F(k\Delta t)}{S_i(k \Delta t)} & \text{ if  } k\bmod m = 0  \\
q^{temp}_i(k\Delta t) & \text{ otherwise }
\end{array}
\right.
\end{align} 

The other method of reducing transaction costs not leaving the idea of balancing the portfolio completely is to build a \emph{partially balanced portfolio}. In this strategy, however, the goal is not to limit the number of transactions, but their value. Therefore, instead of making the portfolio 100\% balanced in every iteration, each time transactions are made that bring the portfolio closer to the state of being fully balanced, but they do not make it entirely balanced. The parameter which will encode what part of portfolio is being rebalanced will be denoted by $D \in (0, 1)$. In each time step and for each asset, instead of buying or selling the amount of assets which would completely balance the portfolio, only a part $D$ is actually exchanged. Consequently, the portfolio is never fully balanced, but the transaction costs are also smaller by a factor of $D$. To conveniently write the formula for updating quantities in this model, we will introduce a new term which we will call $i$-th temporary wealth component $W^{temp}_i(t)$, given by the following formula

\begin{equation}
\label{formula_wealth_component}
W^{temp}_i(k\Delta t) = S_i(k\Delta t)\cdot q^{temp}_i(k\Delta t).
\end{equation} 

This is simply the part of wealth coming from the $i$-th asset. The rebalance in the partial rebalance model is done as per the formula \eqref{formula_updating_quantities_partial_rebalance_fees}

\begin{equation}
\label{formula_updating_quantities_partial_rebalance_fees}
q_i(k\Delta t) = \frac{W^{temp}_i(k\Delta t) + D\cdot (f_i \cdot W^{temp}(k\Delta t) - W^{temp}_i(k\Delta t) - f_i \cdot F(k\Delta t))}{S_i(k\Delta t)}.
\end{equation} 

Note that the formula \eqref{formula_updating_quantities_partial_rebalance_fees} for $D=0$ simplifies to 

\begin{equation*}
q_i(k\Delta t) = \frac{W^{temp}_i(k\Delta t)}{S_i(k\Delta t)} = q^{temp}_i(k\Delta t) = q_i((k-1)\Delta t),
\end{equation*}
which essentially means that no transaction is made and hence -- it reduces to the passive portfolio. For $D=1$ however we obtain  

\begin{equation*}
q_i(k\Delta t) = f_i \frac{W^{temp}(k\Delta t) - F(k\Delta t)}{S_i(k \Delta t)},
\end{equation*}
which is equivalent to the formula for fully balanced portfolio, presented by formula \eqref{formula_updating_quantities_fees}. Hence, for a partially balanced portfolio one could think of $D$ as of a 'slider' between passive and fully balanced portfolio which can be considered two extremes.

\section{Simulation, measurements and results}

In order to study performance of portfolio management strategies described above, we decided to utilise Monte Carlo simulations. We used Geometric Brownian Motion (often abbreviated as GBM) as the model for evolving asset prices. GBM is a well-known stochastic process commonly used for simulation of prices of assets. Admittedly it is a relatively simple model but it has few strong advantages. The most important one is that the stock prices behaviour being described by Geometric Brownian Motion is one of the assumption of the Black-Scholes model, which is arguably the most important model in whole area of financial engineering. From perspective of stochastic calculus, GBM is a process which can be obtained as a solution of the following stochastic differential equation

\begin{equation}
\label{formula_GBM_sde}
\begin{aligned}
dS(t) &= \mu S(t) dt + \sigma S(t) dB(t),\\
S(0) &= s_0,
\end{aligned}
\end{equation}
where $\mu$ and $\sigma$ are parameters of the process, called the drift and the volatility respectively, $s_0$ is the initial value of the process, $B(t)$ is the Wiener process, also known as Brownian motion and $S(t)$ is the GBM itself. 

The solution of equation \eqref{formula_GBM_sde} can be obtained through the It\^{o} formula and is given as follows

\begin{equation}
\label{formula_GBM_explicit}
S(t) = s_0 e^{\left(\mu - \frac{\sigma^2}{2}\right)t + \sigma B(t)}.
\end{equation}

This theoretical result cannot be used directly for the simulation -- it needs to be discretised. To this end, well known defining properties of Brownian motion can be utilised -- starting from the value of $0$, stationarity and independence of its increments and known distribution of these increments. To be precise, for any selection of $s<t$ we have:

\begin{align}
B(0) &= 0, & B(t-s) &\stackrel{d}{=} B(t) - B(s), \nonumber \\
B(s) &\bot B(t) - B(s), & B(t-s) &\sim \mathcal{N}(\mu = 0, \sigma^2 = t-s). \label{formula_Brownian_motion_properties}
\end{align}

From the properties listed in formula \eqref{formula_Brownian_motion_properties} it follows immediately that if we want to simulate a Brownian motion process at the series of points $0, \Delta t, 2\Delta t \ldots$ we first need to simulate a series of normal random variables $Z_1, Z_2, Z_3\ldots$ where for each $k$ $Z_k \sim \mathcal{N}(\mu = 0, \sigma^2 = \Delta t)$ then state $B(0) = 0$ and iteratively repeat 

\begin{equation}
\label{formula_Brownian_motion_simulation}
B(k\Delta t) = B((k-1)\Delta t) + Z_k.
\end{equation}
for each $k = 1, 2, 3 \ldots$. 

Having values of Brownian motion in all desired points, to simulate Geometric Brownian Motion one should use the following discrete version of equation \eqref{formula_GBM_explicit}

\begin{equation}
\label{formula_GBM_simulation}
S(k\Delta t) = s_0 e^{\left(\mu - \frac{\sigma^2}{2}\right)k\Delta t + \sigma B(k\Delta t)}
\end{equation}
for each $k$ for which we need it, i.e. $k \in \mathbb{N}, 1\leqslant k \leqslant \frac{T}{\Delta t}$

To measure performance of proposed portfolios, after Alper et al. \cite{alper_2017} we propose the following quantity, called \emph{growth of wealth}.

\begin{definition}
	\label{definition_growth}
	Growth of portfolio $\mathcal{P}$ -- denoted by $g_{\mathcal{P}}(t)$ -- for any time moment $t \neq 0$ is given by 
	\begin{equation}
	\label{formula_growth}
	g_{\mathcal{P}}(t) = \frac{\log\frac{W(t)}{W(0)}}{t}
	\end{equation}
\end{definition}

The advantages of this measure is that it is not dependent on the initial value of the portfolio (since wealth of portfolio gets divided by its initial value). Imposing the logarithm on the numerator 'flattens' the values of the measure in its right end. Consequently, the change of the value of portfolio becomes relative to its size. Time dependence of an investment has also been captured by the division of the whole amount by $t$. 

\begin{figure}[h]
	\centering
	\includegraphics[width=\textwidth]{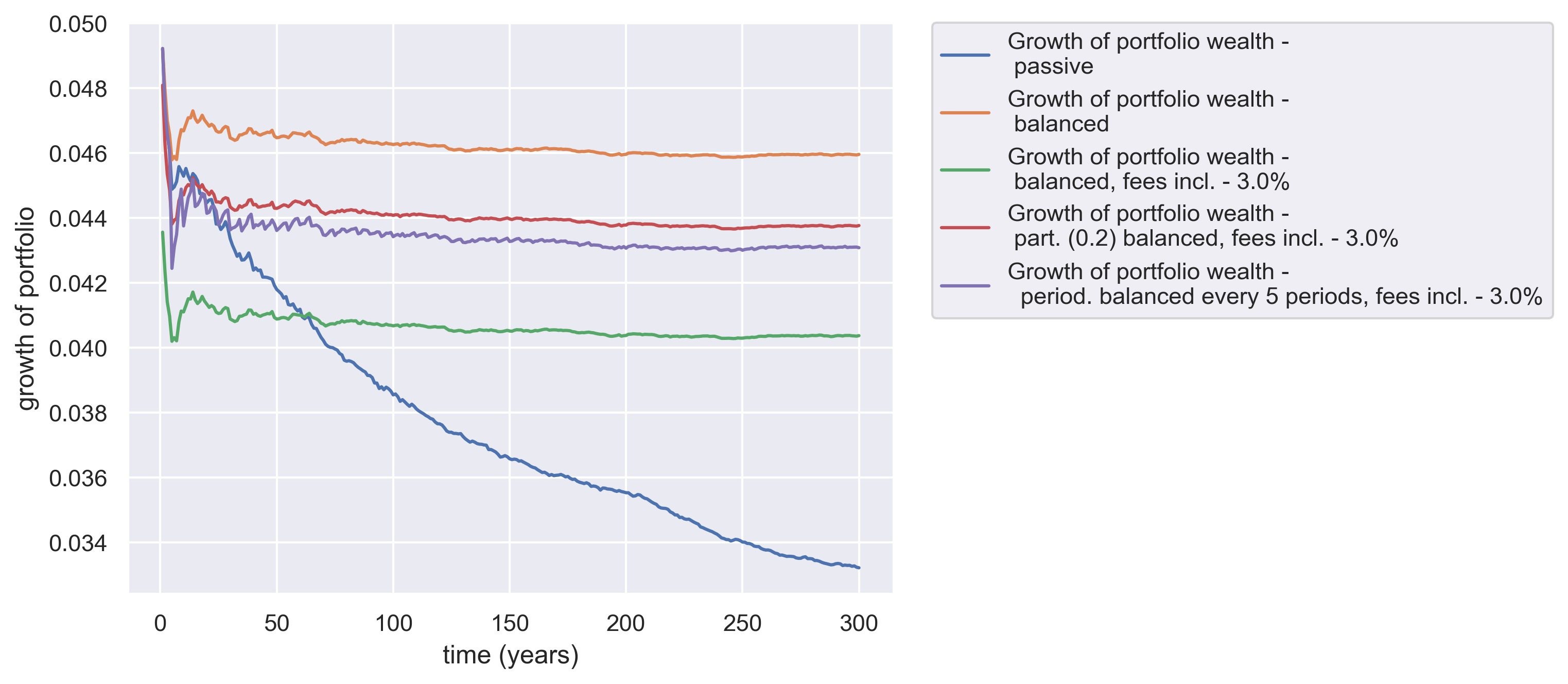}
	\caption[Comparison of various portfolio management startegies]{(Colour online) Portfolio growth in time for various types of portfolios and market conditions. Simulation parameters: $T =300, \Delta t =0.1,  s_0 = 1, \mu = 0.5, \sigma = 1, n=8, MCS=1000$}
	\label{figure_portfolioGrowth_time_variousPortfolios}
\end{figure}

Having the measure of portfolio performance we can proceed to the actual Monte Carlo experiment. To this end one should fix the number of assets in portfolio $n$, simulate trajectories of prices of each of these $n$ assets using formula \eqref{formula_GBM_simulation}, consider different portfolio management strategies and calculate growth of wealth given by the definition \ref{definition_growth} for each of them and last but not least -- repeat it all multiple times, averaging results obtained for different strategies. Results of such experiment are presented in Fig.~\ref{figure_portfolioGrowth_time_variousPortfolios} which shows the average growth of five different portfolios in time. We can easily conclude, that a passive portfolio is the worst performing one. For the costs of transactions equal to $3\%$, the balanced portfolio performs much worse and is only able to outperform the results of a passive portfolio. In order to improve this result, periodic or partial rebalancing strategy can be utilised. In the parameter setup used for Fig.~ \ref{figure_portfolioGrowth_time_variousPortfolios}, both strategies perform quite similarly but partial rebalance seems to be a better choice.

\begin{figure}[h]
	\centering
	\includegraphics[width=\textwidth]{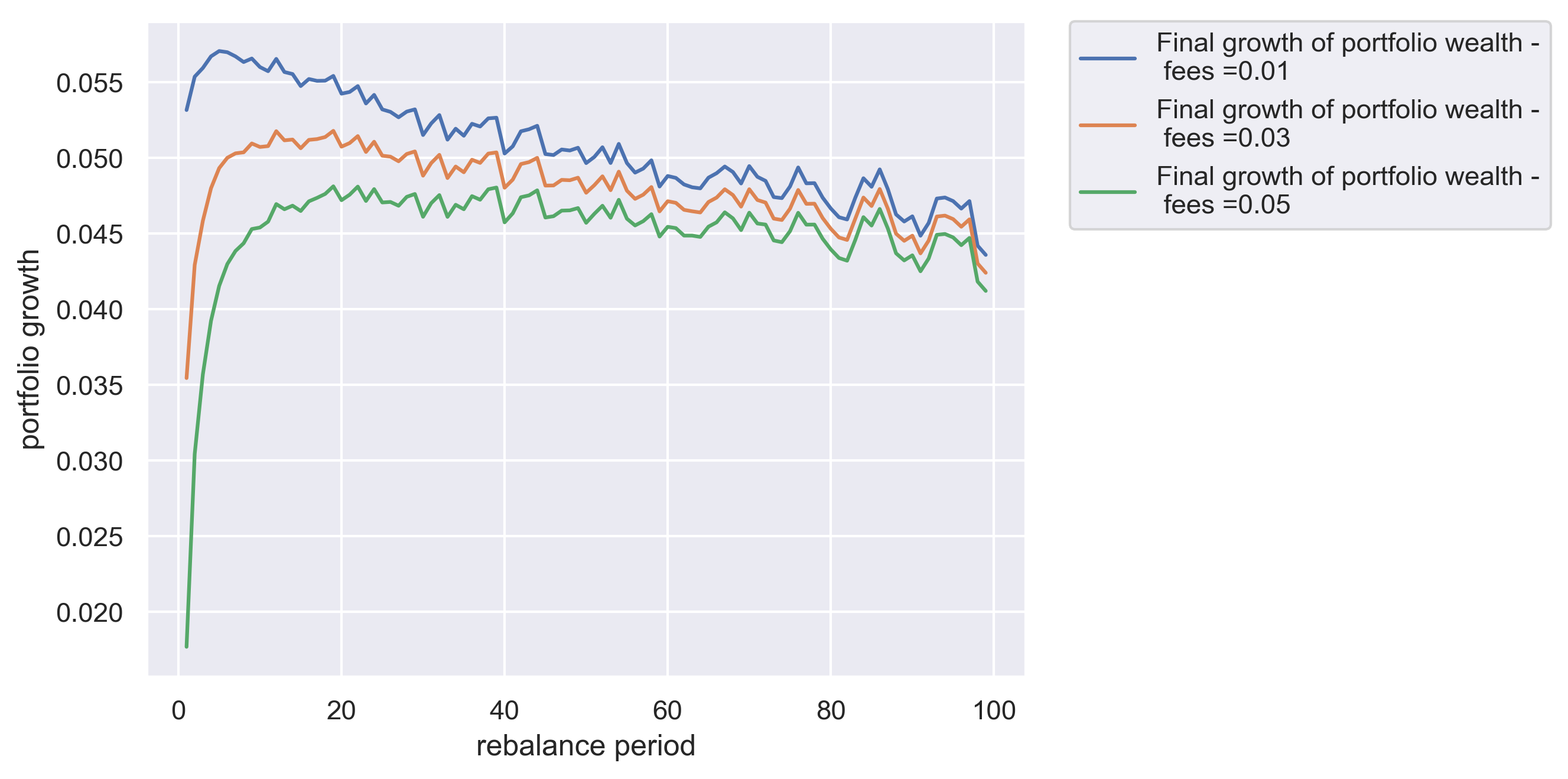}
	\caption[Final growth of periodically balanced portfolio]{(Colour online) Final growth of portfolio (for $t=T$) in dependence of the period of rebalance. A declining trend can be observed, although the variance between results for different rebalance periods is significant. Simulation parameters: $T = 100, \Delta t = 0.1, s_0=1, \mu = 0.125, \sigma = 0.5, n=2, MCS=1000$}
	\label{figure_finalPortfolioGrowth_rebalancePeriod_variousFees}
\end{figure}

\begin{figure}[h]
	\centering
	\includegraphics[width=\textwidth]{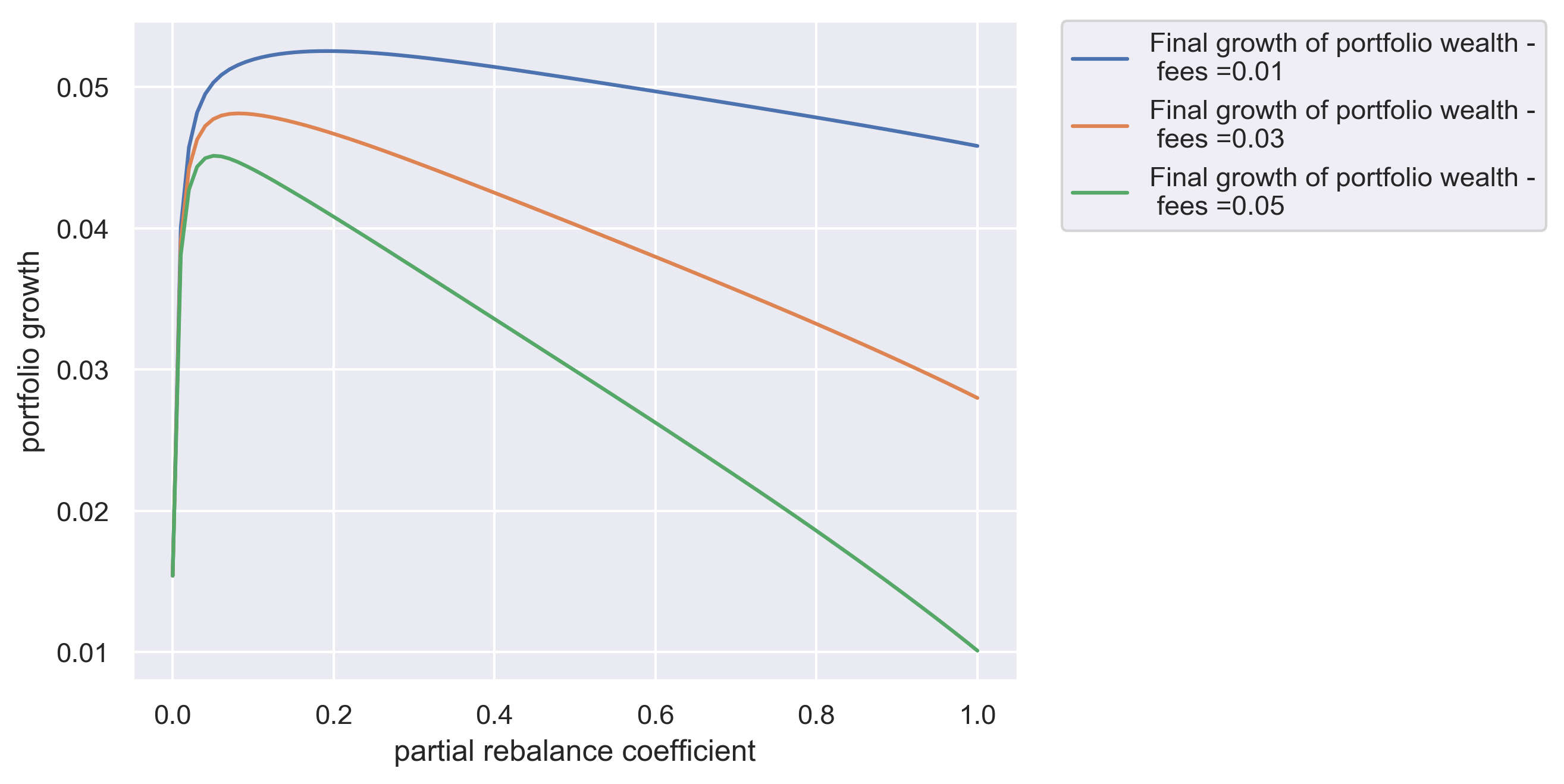}
	\caption[Final growth of partially balanced portfolio]{(Colour online) Final growth of portfolio (for $t=T$) in dependence of the partial rebalance coefficient. In case of bigger fees, the values of the growth significantly drop for coefficients close to $1$, although for smaller fees, the final growth seems to be less dependent on the coefficient. Simulation parameters: $T = 100, \Delta t = 0.1, s_0=1, \mu = 0.125, \sigma = 0.5, n=2, MCS=1000$}
	\label{figure_finalPortfolioGrowth_partialRebalanceCoefficient_variousFees}
\end{figure}

As shown in Fig.\ref{figure_portfolioGrowth_time_variousPortfolios}, partially and periodically balanced portfolios seem to be the best in case of presence of transaction fees. We decided to study what parameters of these portfolios maximise the growth of wealth of portfolio at the point where simulation terminates (i.e. at portfolio's maturity time $T$). The level of transactions fees $\alpha$ turned out to have crucial meaning when determining the optimal parameters. For periodically balanced portfolios, the bigger the fees, the more rarely rebalancing should be done (the value of the multiplier $m$ should be bigger) and for partially balanced portfolios for higher fees portfolios should be rebalanced in smaller scope (value of parameter $D$ should be smaller). These observations have been visualised in Figs. \ref{figure_finalPortfolioGrowth_rebalancePeriod_variousFees} and \ref{figure_finalPortfolioGrowth_partialRebalanceCoefficient_variousFees}.

\begin{figure}[h]
	\centering
	\includegraphics[width=\textwidth]{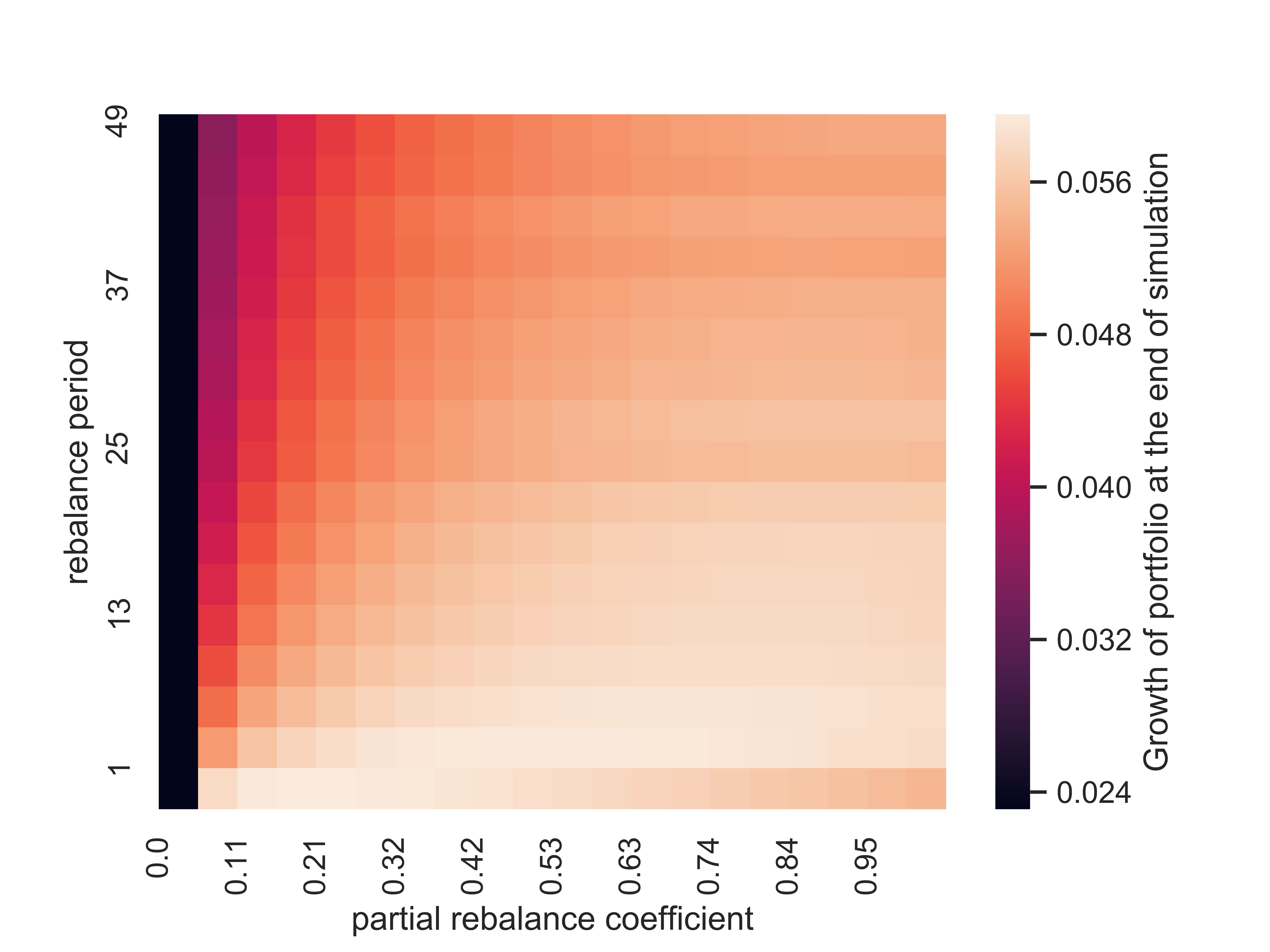}
	\caption[Final growth of periodically and partially balanced portfolio]{(Colour online) Final growth of portfolio (for $t=T$) in dependence of the period of rebalance and partial rebalance coefficient. It can be observed that for bigger rebalance periods, bigger partial rebalance coefficients are needed to receive better results. Leftmost column, corresponding to the partial rebalance coefficient equal to $0$ is equivalent to passive portfolios, which exhibits poorest results. Simulation parameters: $T = 100, \Delta t = 0.1, s_0=1, \mu = 0.125, \sigma = 0.5, n=2, F = 0.01, MCS=1000$}
	\label{figure_partialRebalanceCoefficient_rebalancePeriod_heatmap}
\end{figure}

Periodically and partially balanced portfolios can also be easily merged together and hence -- the resulting strategy should be dependent on both period of rebalance $m$ and partial rebalance coefficient $D$. To visualise the dependence of the growth of wealth on both of these parameters, we created a heatmap, which is presented in Fig.  \ref{figure_partialRebalanceCoefficient_rebalancePeriod_heatmap}. It can be clearly seen that for partial rebalance coefficient equal to 0 (leftmost part of the map) results are very poor. As noted before, setting this parameter to 0 results in reduction to passive portfolio which returns poor results. However, for any other value of partial rebalance coefficient, results improve significantly. The map also confirms that for frequent rebalancig, it's best to perform smaller rebalancing operations and if we rebalance portfolio more rarely -- bigger rebalancing operations are advantageous. Also -- the more one rarefies the moments of the rebalance (longer rebalance period) the worse results they get (upper part of the map is generally much darker than the lower one). 

Analysing performance of a portfolio based on simulated data may rise objections since for the Monte Carlo simulations one needs to pick a way to model the price dynamics of the assets. A~model, on the other hand, serves only as an approximation of the reality and never describes it perfectly. Therefore we undertook a number of numerical experiments on real, stock exchange market data in order to see if our findings, also apply to the real-world portfolios.

The data for an experiment comes from Warsaw Stock Exchange (\emph{Giełda Papierów Warto\'{s}ciowych w Warszawie} in Polish) and it involves the prices of shares of companies grouped into WIG20 index. WIG20 consists of twenty biggest Polish companies by market capitalization and hence -- it is one of the most important indicators of Polish economy. The data covers daily prices of shares of those twenty companies from the time range between 2nd January 2015 and 31st March 2018. The data has been downloaded from Stooq.pl \footnote{Stooq.pl is a webside providing data from Polish and international stock exchange markets. It is particularly popular among Polish investors.}.

To clearly see the difference between the results obtained with simulated and real data, it would be convenient to repeat the analysis that were made for simulated portfolios, simply changing method of creating them. This is not possible directly however due to the fact that using pure Monte Carlo simulations one is able to create arbitrary number of different portfolios which can later be averaged but having the limited amount of data (in our case - finite number of assets to choose from), this cannot be done. However, in order not to rely on the result of just one or two portfolios constructed entirely by conscious selection (like it was done by other authors, e.g. \cite{alper_2017}), we developed a method which is supposed to mimic (to some extent) the Monte Carlo experiment, but with limited amount of data. First, we need to fix the number of assets inside one portfolio $n$ and the number of all assets available on the market -- $N$. For us, $N=20$, therefore the number of all possible portfolios that can be constructed (in terms of the scope of assets) is the number of combinations of $n$-element subsets of an $N$-element set. This number can be calculated as $\binom{N}{n} = \binom{20}{n}$. For example, if we consider the set of 2-assets portfolios, we can make $190$ of them, and for 8-assets portfolios -- $125,970$ portfolio compositions are possible. Hence -- the only thing that needs to be fixed is the number of stocks in portfolio $n$ and the number of portfolios among all $\binom{N}{n}$ that will be randomly drawn to appear as part of an experiment. Finally, the average result is calculated from the results of randomly chosen portfolios. Such value can be viewed as providing more information about the general behaviour of the assets than the result from a single portfolio.

\begin{figure}[h]
	\centering
	\includegraphics[width=\textwidth]{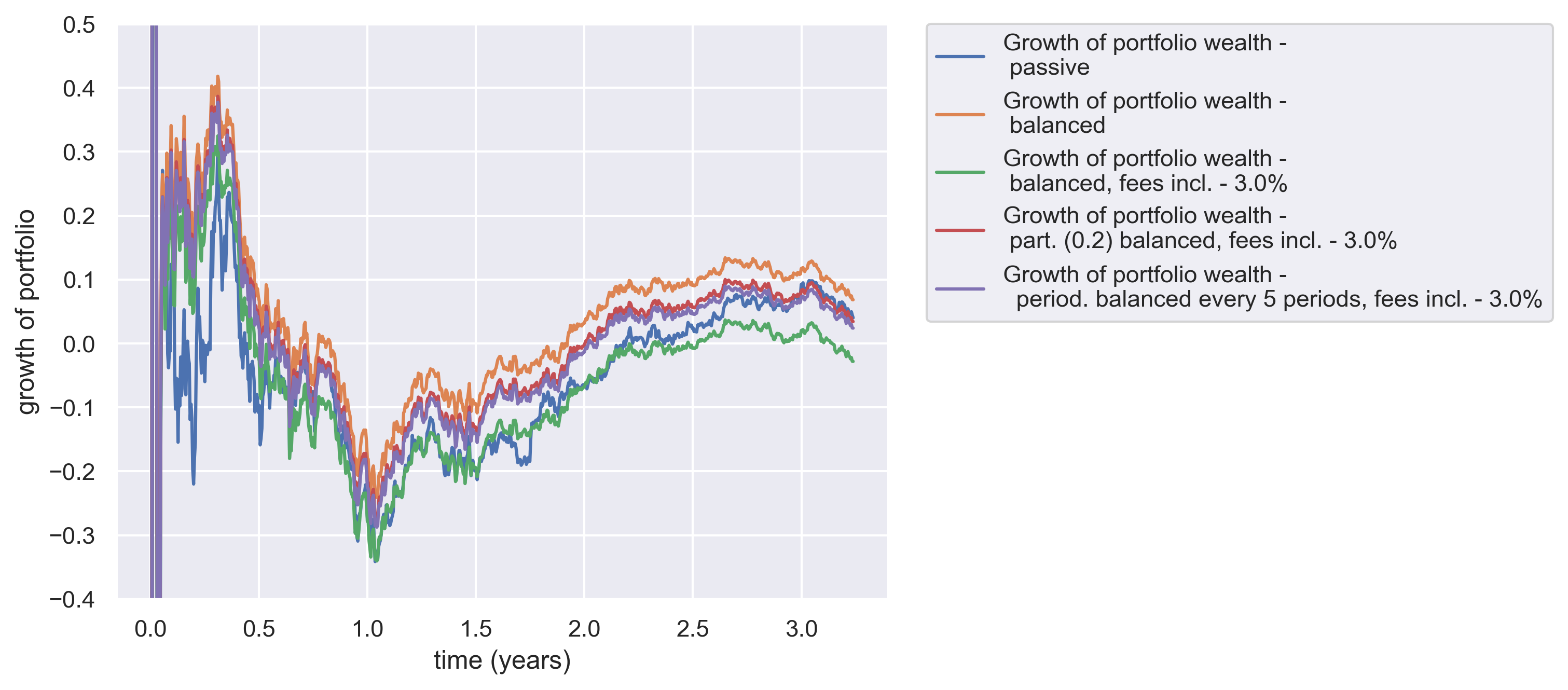}
	\caption[Various startegies for market data portfolios]{(Colour online) Portfolio growth in time for various types of real-data portfolios and market conditions, averaged over 1000 runs. For each iteration a new 8-element subset of 20 available stocks was picked to build each portfolio.}
	\label{figure_marketData_portfolioGrowth_time_variousPortfolios_biggerFees}
\end{figure}

From Fig.~\ref{figure_marketData_portfolioGrowth_time_variousPortfolios_biggerFees} we can see that the differences between the growth of rebalance-related portfolios seems not to significantly outperform the passive one. This suggests that, although our predictions based on the GBM simulations somewhat find the confirmation in the real-data experiment, they also turned out to overrate the rebalance-related methods compared to the simple, passive one. We found that the main factor impacting the prepotency of the more complicated portfolio management strategies are the fees coefficient and the time of the investment.

\begin{figure}[h]
	\centering
	\includegraphics[width=\textwidth]{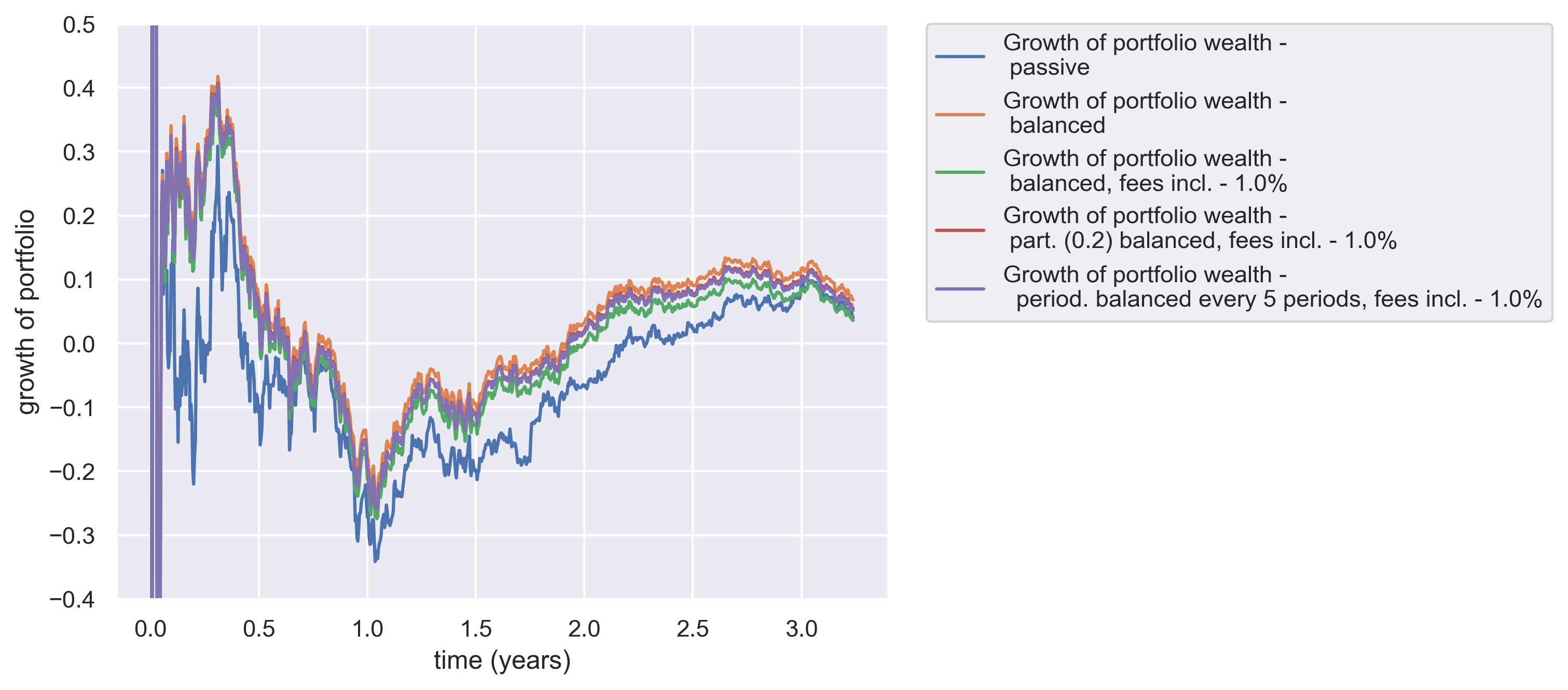}
	\caption[Various startegies for market data portfolios]{(Colour online) Portfolio growth in time for various types of real-data portfolios and market conditions, averaged over 1000 runs. For each iteration a new 8-element subset of 20 available stocks was picked to build each portfolio.}
	\label{figure_marketData_portfolioGrowth_time_variousPortfolios_smallerFees}
\end{figure}

Fig.~\ref{figure_marketData_portfolioGrowth_time_variousPortfolios_smallerFees} presents the results of the same experiment as Fig.~\ref{figure_marketData_portfolioGrowth_time_variousPortfolios_biggerFees} but with fees level decreased from 3\% to 1\%. Curves representing periodical and partial rebalance strategies are noticeably shifted upwards. Also, the fully-balanced portfolio presents much bigger values of the growth here. But even for the lower level of fees, the gap between average passive portfolio and average balanced ones (including periodically or partly balanced portfolios) is far less impressive compared to the results of our pure-Monte-Carlo simulations (see Fig.~\ref{figure_portfolioGrowth_time_variousPortfolios}).

One can notice that for all strategies that have been presented in Figs.~\ref{figure_marketData_portfolioGrowth_time_variousPortfolios_smallerFees} and~\ref{figure_marketData_portfolioGrowth_time_variousPortfolios_biggerFees} a deep minimum of the value of portfolio growth can be seen around the time point 1. This is however not due to a failure of the strategies themselves -- it turns out that January 2016 (which is more or less equivalent to the point 1) was very ‘bearish’ for Polish stock market –- it started a recession period which lasted for a year approximately. This very recession is reflected in small values of the portfolio growth in a time interval close to 1. 

\begin{figure}[h]
	\centering
	\includegraphics[width=\textwidth]{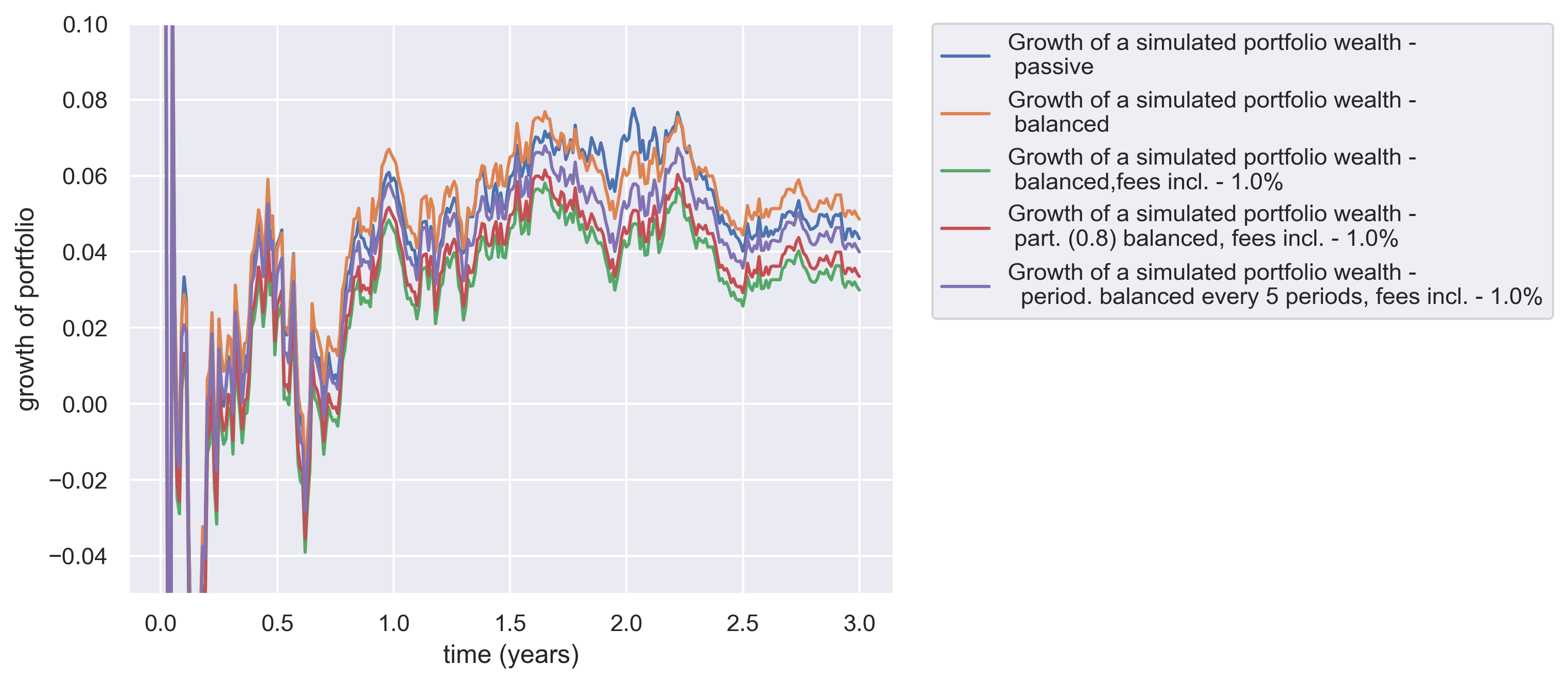}
	\caption[Various startegies for simulated portfolios drawn from finite set with $T=3$]{(Colour online) Portfolio growth in time for various types of portfolios and market conditions when the content of portfolios was being drawn from a finite set of assets, all averaged over 1000 runs. For each iteration a new 8-element subset of 20 available stocks was picked to build each portfolio. Simulation parameters: $T = 3, \Delta t = 0.1, s_0=1, \mu = 0.125, \sigma = 0.5, n=2, F = 0.01, MCS=1000$}
	\label{figure_drawingFromSimulatedData_portfolioGrowth_time_variousPortfolios_T3}
\end{figure}

\begin{figure}[h]
	\centering
	\includegraphics[width=\textwidth]{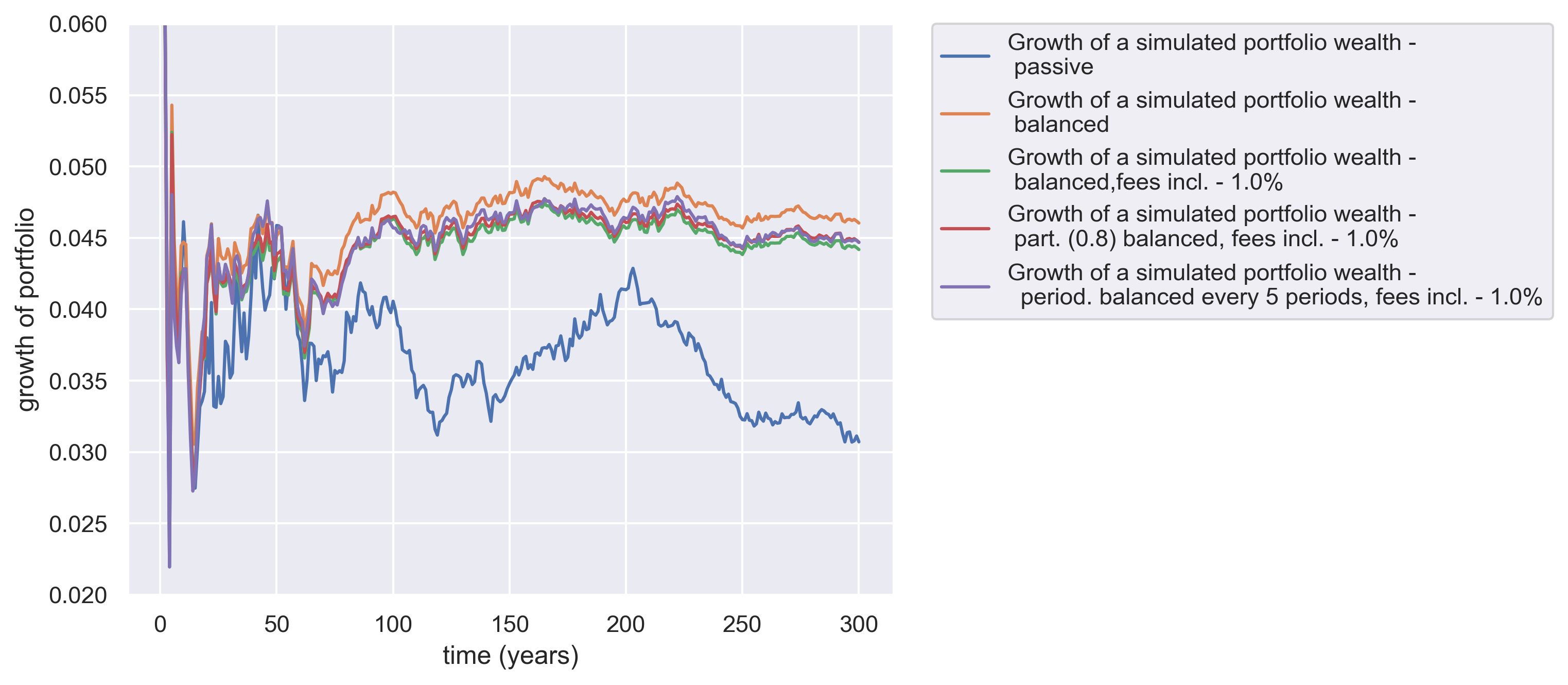}
	\caption[Various startegies for simulated portfolios drawn from finite set with $T=300$]{(Colour online) Portfolio growth in time for various types of portfolios and market conditions when the content of portfolios was being drawn from a finite set of assets, all averaged over 1000 runs. For each iteration a new 8-element subset of 20 available stocks was picked to build each portfolio. Simulation parameters: $T = 300, \Delta t = 0.1, s_0=1, \mu = 0.125, \sigma = 0.5, n=2, F = 0.01, MCS=1000$}
	\label{figure_drawingFromSimulatedData_portfolioGrowth_time_variousPortfolios_T300}
\end{figure}

We conducted a dedicated experiment to explain why the difference between the rebalance-related portfolios and a simple passive portfolio is a lot bigger in Monte Carlo simulations than for the market-data portfolios. In the experiment we simulated two sets of data -- for the first one we simulated prices of 20 assets over the course of 3 years and for the other one we simulated prices of 20 assets again, but this time over the course of 300 years (similar maturities were also studied in the paper of Alper et al. \cite{alper_2017}). Then for both datasets we employed a similar technique of constructing portfolios as when dealing with our true market data ones (choosing them from a finite set of possibilities). It turned out that the results for the 3-year portfolios, presented in the Fig. \ref{figure_drawingFromSimulatedData_portfolioGrowth_time_variousPortfolios_T3} look very similarly to ones which we constructed using market data (to recall - Fig. \ref{figure_marketData_portfolioGrowth_time_variousPortfolios_smallerFees}) in a sense that passive portfolio does not seem to perform much worse than the rebalance-related ones. However Fig. \ref{figure_drawingFromSimulatedData_portfolioGrowth_time_variousPortfolios_T300} which shows the same portfolios, but with maturity equal to $300$ years proves that the passive portfolio is comparable to the other ones only at the very beginning of portfolio's lifetime and as the time passes -- differences between the passive portfolio and all the remaining ones become evident. It is also notable that despite we used simulated data in this experiment (for both $3-$ and $300-$year portfolio), the lines associated with portfolios are now much more ragged and indented. We can conclude that this is an effect of the method of constructing portfolios, i.e. building them using only finite number of assets, as described previously. Only in case of Monte Carlo simulations, with a possibility of generating arbitrary numbers of asset trajectories one is able to observe the perfectly smoothened average growths of portfolios as a result.

\section{Conclusions}

In the paper various portfolio management strategies were presented, both from theoretical and application side. Each strategy was  described in detail and a precise and rigorous algorithm of their usage was given. We have confirmed earlier research claiming that certain portfolios -- namely the partially and periodically rebalance ones -- are better than the others in case of the presence of transaction fees. Transaction fees were actually shown to be the crucial aspect which should be strongly considered while planning the portfolio management strategy. We showed, based on the Monte Carlo simulations and assuming asset price dynamic given by Geometric Brownian Motion, that it is possible to optimise partial rebalance coefficient and periodic rebalance term to maximise growth of portfolio. We have also conducted an experiment in which we merged together partial and periodic rebalancing to create a new strategy dependent on the parameters from both of these methods and we have shown how the growth of wealth looks like for such a portfolio construct, depending on parameters taken from each of the models. To compare results from Monte Carlo simulations to the the ones of the actual market data, we devised a specific method of testing portfolio management strategies when the set of available asset trajectories is finite. We managed to show that for portfolios created in such manner the behaviour changes significantly compared to classical Monte Carlo approach. However, we have also showed that the same methodology applied to the simulated trajectories of assets gives similar results. Finally, we also managed to prove that the difference between classical Monte Carlo and our approach is significantly lessened for long-lasting portfolios. Surely -- more research is needed to answer multiple questions that emerged throughout the article, e.g. what is the impact of the drift and variability parameters for the results of our portfolios? In what way the underlying model (GBM) impacts the simulation? Are there any other analytically describable strategies worth considering? The answer to these questions may contribute a lot to the general research on the topic of mathematical modelling of investment portfolios, which is both fascinating and nondepletable topic. 

\section*{Acknowledgement}

This work was supported by the Polish Ministry of Science and Higher Education (MNiSW) core funding for statutory R\&D activities.

\bibliographystyle{elsarticle-num}
\bibliography{Bibliography}
\end{document}